\documentclass{article}   
\usepackage{amsmath}       
\DeclareMathOperator{\arcsinh}{arcsinh}         \allowdisplaybreaks		
\usepackage{hyperref}

\title{Considering the two Spin and the two Angular Momenta String Solutions in $AdS_5 \times S^5$}	
\author{Arne L. Larsen, \href{mailto:all@sdu.dk}{all@sdu.dk}\\ University of Southern Denmark and FKF\\ Also at cp3}
\date{}
\begin{document}
\maketitle
\begin{center}
ABSTRACTS
\end{center}
In this paper, we consider two almost opposite sectors of actual string configuration ansätze in $AdS_5\times S^5$, which anyway have almost the same features: The two spin solution, which has constant angles in $S^5$ and the two angular momenta solution, which has constant "angels" in $AdS_5$, however for the two angular momenta solution, we have to take the time coordinate from $AdS_5$, thus there is a little asymmetry between the two string configurations in $AdS_5\times S^5$. \\
\indent Without being autistic, there is around $69$ equations between the similar equations in the two sectors, compare equations (34) and (104) and also compare equations (64) and (134). Again, without being autistic, the text after the equations in the two sectors, is almost precisely the same. In our notation, the difference between the two sectors is as follows; $\rho\leftrightarrow \theta$, $\phi \leftrightarrow \psi$, $\sinh\rho \leftrightarrow  \sin\theta$, $y_i\leftrightarrow x_i$, $y \leftrightarrow x$ etc. \\
\indent The string configurations of this paper, are both solvable by the Neumann System. However, our setup in this paper is generally for the Neumann-Rosochatius System, which is also solvable, since we intend to generalize our results from the Neumann System to the Neumann-Rosochatius System and to several types of deformed Neumann-Rosochatius Systems. \\
\indent In the second part of this paper, which is independent of any string configurations in $AdS_5\times S^5$ and concerns String Cosmology in $D=10$ dimensions. I will seriously argue that there was no Big Bang; I truly believe that the Universe has been there forever, see True Conclusions, Section 10.\footnote{For more than twenty years, I have had ALS, as Stephen Hawking had, so today, I can only move my eyeballs. In addition, I have a ventilator, thus I cannot breathe myself. I could not breathe myself since New Years eve 2005/2006, that is more than sixteen years ago, and Happy New Year....}
\newpage
\section{Introduction}
In some sense, one can argue that the AdS/CFT correspondence started by the logarithmic dependence in $E(S)$ of the open spinning string in $AdS_3$, for long strings, in the paper of de Vega and Egusquiza \cite{deVega:1996mv}. Here $E$ is the energy and $S$ is the spin of the string configuration. Furthermore, the logarithmic dependency in $E(S)$ for long strings, is very typical in the AdS/CFT correspondence. The logarithmic dependence is also present for the string solutions in the present paper \cite{Larsen:2003tb}, because of unconventionally boundary conditions, see below and see also \cite{Khan:2003sm}. For another notation, see also \cite{Tirziu:2009ed}.  \\\\
The action for a string in curved spacetime is, in the conformal gauge
\begin{equation}
I=-\frac{\sqrt{\lambda}}{4\pi}\int d\tau d\sigma g_{\mu\nu}(X)\partial_AX^\mu\partial^AX^\nu, \qquad \sqrt{\lambda}\equiv\frac{R^2}{\alpha^\prime}
\end{equation}
The string equations of motion, in the conformal gauge are 
\begin{equation}
\ddot{X}^\mu-X^{\prime\prime\mu}+\Gamma^{\mu}_{\rho\sigma}(\dot{X}^{\rho}\dot{X}^\sigma-X^{\prime\rho}X^{\prime\sigma})=0 
\end{equation}
\begin{equation}
g_{\mu\nu}\dot{X}^\mu X^{\prime\nu}=0, \quad \quad g_{\mu\nu}(\dot{X}^{\mu}\dot{X}^\nu+X^{\prime\mu}X^{\prime\nu})=0 
\end{equation}
where dot and prime denote derivative with respect to $\tau$ and $\sigma$, respectively.\\\\
We now consider actual string configurations in $AdS_5$, $S^5$ and in $AdS_5\times S^5$. Then the pioneering papers were \cite{Frolov:2003qc, Arutyunov:2003uj, Arutyunov:2003za, Tseytlin:2003ii, Frolov:2002av, Tseytlin:2002ny, Frolov:2003tu, Frolov:2003xy}, \cite{Larsen:2003tb, Khan:2003sm,Khan:2005fc, Ryang:2004tq, Larsen:2007va} and \cite{deVega:1996mv, deVega:1993hq, deVega:1994yz, Larsen:1994ah, Larsen:1996gn, Larsen:1995bp, deVega:1993rm, Larsen:1994mr} and references therein. Notice that all the papers from the Paris group \cite{deVega:1996mv, deVega:1993hq, deVega:1994yz, Larsen:1994ah, Larsen:1996gn, Larsen:1995bp, deVega:1993rm, Larsen:1994mr}, were published long before the AdS/CFT correspondence \cite{Maldacena:1997re, Gubser:1998bc, Witten:1998qj}, see also \cite{Minahan:2002ve, 2004NuPhB.676....3B, Beisert:2003yb, Beisert:2003tq}. \\\\
The line element of $AdS_5\times S^5$ is \\
\begin{align}
ds^2=&-\cosh^2\rho dt^2 + d\rho^2 + \sinh^2\rho (d\phi^2 +\cos^2\phi d\phi_1^2+\sin^2\phi d\phi_2^2)\notag\\
& +d \theta^2+\cos^2 \theta d \psi_3^2+\sin^2 \theta(d \psi^2 +\cos^2\psi d \psi_1^2 +\sin^2 \psi d \psi_2^2)
\end{align}
Let us go back to 2003. Frolov and Tseytlin \cite{Frolov:2003qc} first "proved" that string solutions could not depend on the radial coordinate $\rho$ and $\theta$, respectively in the two sectors, in our notation, but they were wrong! Later this year, a student and I, Alex Khan and I, found the new multi spin solution, which existence was due to some unconventionally boundary conditions: \\\\
On the 'radial' coordinate $\sinh\rho$, we impose the periodicity condition
\begin{equation}
\sinh\rho(\sigma+2\pi) = \sinh\rho(\sigma)\label{new:17}
\end{equation}
On the angular coordinate, we impose the quasi periodicity condition
\begin{equation}
\phi(\sigma + 2\pi) = \phi(\sigma) + 2M\pi\label{12}
\end{equation}
where $M$ is a positive integer.\\\\
However, equation \eqref{new:17} must be refined. The boundary condition will eventually be imposed on $\sinh^2\rho$ and we shall consider a $2N$ ($N$ positive integer) folded string in the radial direction, thus we impose the boundary condition
\begin{equation}
\sinh^2\rho(\sigma+\frac{2\pi}{2N})=\sinh^2(\sigma)\label{13}
\end{equation}
This does not mean that the actual string is folded; it merely consists of $N$ 'arcs'.\\\\
Notice that Frolov and Tseytlin and others, already during 2003, found out that they were wrong \cite{Arutyunov:2003uj, Arutyunov:2003za, Tseytlin:2003ii, Tseytlin:2002ny, Frolov:2003tu, Frolov:2003xy}. \\\\
The string configurations of this paper, are both solvable by the Neumann System. However, our setup in this paper is generally for the Neumann-Rosochatius System, which is also solvable, since we intend to generalize our results from the Neumann System to the Neumann-Rosochatius System \cite{Arutyunov:2003za, ADLER1980318, 10.1007/978-1-4613-8109-9_7, Ratiu1981EulerPoissonEO, Ratiu1981TheCN, Reiman82, RosochatiusPHD}, see also \cite{Chakraborty:2019gmt, Ahn:2008hj, Hernandez:2015nba, Chakraborty:2020las} etc., and to several types of deformed Neumann-Rosochatius Systems \cite{Hernandez:2017raj, Arutyunov:2016ysi, Nieto:2017wtp} etc.\\\\
This paper is based solely on \cite{Larsen:2003tb}, for another notation, see \cite{Tirziu:2009ed}, but I also want to cite the following authors for excellent literature: Lewis Carroll \cite{Caroll1865}, Francis Scott Fitzgerald \cite{Fitzgerald1925} and the Nobel Prize winner in Literature Henrik Pontoppidan \cite{Pontoppidan1898}.  

\section{The open Spinning String of de Vega and Egusquiza}

The action for a string in curved spacetime is, in the conformal gauge
\begin{equation}
I=-\frac{\sqrt{\lambda}}{4\pi}\int d\tau d\sigma g_{\mu\nu}(X)\partial_AX^\mu\partial^AX^\nu, \qquad \sqrt{\lambda}\equiv\frac{R^2}{\alpha^\prime}\label{1}
\end{equation}
In some sense, one can argue that the AdS/CFT correspondence started by the logarithmic dependence in $E(S)$ of the open spinning string in $AdS_3$, for long strings, in the paper of de Vega and Egusquiza \cite{deVega:1996mv}. Here $E$ is the energy and $S$ is the spin of the string configuration. Furthermore, the logarithmic dependency in $E(S)$ for long strings, is very typical in the AdS/CFT correspondence.

We now review the open spinning string of de Vega and Egusquiza in $AdS_3$, in a modern formulation. 
\\\\
The line element of $AdS_3$ is

\begin{equation}
\text{d}s^2=-\cosh^2\rho \ \text{d}t^2+\text{d}\rho^2+\sinh^2\rho \;\text{d}\phi^2.\label{2}
\end{equation}
Under the ansatz 
\begin{align}
t=\kappa\tau,\phi=\omega\tau,\rho=\rho(\sigma)\label{3}
\end{align}

The string equations of motion, in the conformal gauge are 
\begin{equation}
\ddot{X}^\mu-X^{\prime\prime\mu}+\Gamma^{\mu}_{\rho\sigma}(\dot{X}^{\rho}\dot{X}^\sigma-X^{\prime\rho}X^{\prime\sigma})=0 \label{new:15}
\end{equation}
\begin{equation}
g_{\mu\nu}\dot{X}^\mu X^{\prime\nu}=0, \quad \quad g_{\mu\nu}(\dot{X}^{\mu}\dot{X}^\nu+X^{\prime\mu}X^{\prime\nu})=0 \label{new:16}
\end{equation}
where dot and prime denote derivative with respect to $\tau$ and $\sigma$, respectively.
Equation (\ref{new:15}) and equation (\ref{new:16}) lead consistently  to
\begin{equation}
 \rho^{\prime 2}  = \kappa^2\cosh^2 \rho-\omega^2 \sinh^2\rho \label{ads3-2}
\end{equation} 
The endpoints of the open string, is obtained when $\rho^\prime=0$, which leads to
\begin{equation}
\rho_\pm=\pm\arcsinh\frac{\kappa}{\sqrt{\omega^2-\kappa^2}}
\end{equation}
The explicit solution of the norm of the radial coordinate, can be found as follows.
We substitute
\begin{equation}
y=\sinh^2\rho \label{4}
\end{equation}
then we get
\begin{equation}
y^{\prime2}=4y(1+y)(\kappa^2(y+1)-\omega^2y) \label{yprime}
\end{equation}
the solution is 
\begin{equation}
y^2(\sigma)=\frac{\kappa^2}{\omega^2-\kappa^2}\left(1-\text{sn}^2\left(\omega\sigma\left.\right|\frac{\kappa^2}{\omega^2}\right)\right)
\end{equation}
We only consider $ \omega > \kappa$, otherwise there are no endpoints of the string. Eventually, we will be interested in  long strings, so we will take the limit $\omega \approx \kappa$ (but still $\omega>\kappa$). The energy and the spin of this string solution are \cite{Khan:2003sm}

\begin{align}
E&=\frac{2\omega\kappa}{\pi\alpha^\prime(\omega^2-\kappa^2)}E\left(\frac{\kappa^2}{\omega^2}\right)\\
S&=\frac{2}{\pi\alpha^\prime}\left(\frac{\omega^2}{\omega^2-\kappa^2}E\left(\frac{\kappa^2}{\omega^2}\right)- K\left(\frac{\kappa^2}{\omega^2}\right)\right)
\end{align}

\noindent We now consider strings in the long string limit $\omega\approx \kappa$ (but still $\omega>\kappa$) \cite{Khan:2003sm}
\begin{align}
E - S &\approx \frac{1}{\pi\alpha^\prime}\log\left(\frac{\pi \alpha^\prime S}{2}\right)
\end{align}
This is the leading Regge trajectory. In the next section, we wii embed this open string solution, as a closed string solution in $AdS_5\times S^5$. 


\section{The two Spin String Solution in $AdS_5\times S^5$}
The two spin string solution was originally found in $AdS_5$ \cite{Larsen:2003tb}, so we take constant angels in $S^5$. \\\\ 
The line element of $AdS_5\times S^5$ is \\
\begin{align}
ds^2=&-\cosh^2\rho dt^2 + d\rho^2 + \sinh^2\rho (d\phi^2 +\cos^2\phi d\phi_1^2+\sin^2\phi d\phi_2^2)\notag\\
& +d \theta^2+\cos^2 \theta d \psi_3^2+\sin^2 \theta(d \psi^2 +\cos^2\psi d \psi_1^2 +\sin^2 \psi d \psi_2^2)
\end{align}

\noindent The ansatz for the string solution in $AdS_5\times S^5$ is \begin{align}
&t=\kappa\tau+\beta_3(\sigma)\text{, }\rho=\rho(\sigma)\text{, } \notag\phi=\phi(\sigma)\text{, }\phi_1=\omega_1\tau+\beta_1(\sigma)\text{, }\phi_2=\omega_2\tau+\beta_2(\sigma)\text{, }\\& \theta=\theta_0\text{, } \psi=\psi_0\text{, } \psi_1=\psi_{1_0}\text{, } \psi_2=\psi_{2_0}\text{, }\psi_3 = \psi_{3_0}\label{eq:108}
\end{align}
where $\theta_0$, $\psi_0$, $\psi_{1_0}$, $\psi_{2_0}$ and $\psi_{3_0}$ are constants.\\\\
We introduce the embedding coordinates ($y_1,y_2,y_3$)
\begin{equation}
y_1=\sinh\rho\cos\phi,\quad y_2=\sinh\rho\sin\phi, \quad y_3 = \cosh\rho 
\end{equation} 
They fulfill the Neumann-Rosochatius  integrable system for the metric
\begin{equation}
\eta^{ij} = 
\begin{pmatrix}
-1 & 0 & 0\\
0 & -1 & 0\\
0 & 0 & +1
\end{pmatrix}
\end{equation}
i.e.
\begin{align}
\eta^{ij}y_iy_j\equiv-y_1^2-y_2^2+y_3^2=1\label{eq:71}
\end{align}
We have the periodicity conditions 
\begin{align}
y_i(\sigma+2\pi)=y_i(\sigma)\text{, }\quad \beta_i(\sigma+2\pi)=\beta(\sigma)+2\pi k_i,
\end{align}
where $k_i$ are integers. To have $t$ single valued, we must choose $k_3=0$.\\\\
The equations of motion determine that
\begin{align}
&\beta_3^{\prime\prime}+2\tanh\rho\beta_3^{\prime}\rho^{\prime}=0\label{eq:103},\\\notag\\
&\rho^{\prime\prime}=\cosh\rho\sinh\rho(\kappa^2-\beta_3^{\prime 2})+\phi^{\prime 2}\sinh\rho\cosh\rho\notag\\&-\cos^2\phi\cosh\rho\sinh\rho(\omega_1^2-\beta_1^{\prime 2})\notag\\&-\sin^2\phi\cosh\rho\sinh\rho(\omega_2^2-\beta_2^{\prime 2}),\\\notag\\&\phi^{\prime\prime}=\cos\phi\sin\phi(\omega_1^2-\beta_1^{\prime 2})-\sin\phi\cos\phi(\omega_2^2-\beta_2^{\prime 2})-2\phi^{\prime}\rho^{\prime}\coth\rho,\\\notag\\ &\beta_1^{\prime\prime}-2\phi^{\prime}\beta_1^{\prime}\tan\phi+2\rho^{\prime}\beta_1^{\prime} \coth\rho=0,\label{eq:104}\\\notag\\ &\beta_2^{\prime\prime}+2\phi^{\prime}\beta_2^{\prime}\cot\phi+2\rho^{\prime}\beta_2^{\prime}\coth\rho=0,\label{new:18}
\end{align}
while the constraints give
\begin{align}
&\omega_1\beta^\prime_1\cos^2\phi\sinh^2\rho+\omega_2\beta^\prime_2\sin^2\phi\sinh^2\rho=\kappa\beta^\prime_3\cosh^2\rho,\\\notag\\
\rho^{\prime 2} =&\cosh^2\rho(\kappa^2+\beta_3^{\prime 2})-\phi^{\prime 2}\sinh^2\rho- \sinh^2\rho\cos^2\phi(\omega_1^2+\beta_1^{\prime 2})\notag\\&-\sinh^2\rho\sin^2\phi(\omega_2^2+\beta_2^{\prime 2}).
\end{align}
Equations (\ref{eq:103}), (\ref{eq:104}) and (\ref{new:18}) are trivially solved by
\begin{align}
\beta_i^\prime=\frac{u_i}{y_i^2},\label{new:1}
\end{align}
where $u_i$ are constants. Then the remaining equations of motion lead to
\begin{align}
&\rho^{\prime\prime}=\cosh\rho\sinh\rho(\kappa^2-\frac{u_3^2}{\cosh^4\rho})+\phi^{\prime 2}\sinh\rho\cosh\rho\notag\\&-\cos^2\phi\cosh\rho\sinh\rho(\omega_1^2-\frac{u_1^2}{\sinh^4\rho\cos^4\phi})\notag\\&-\sin^2\phi\cosh\rho\sinh\rho(\omega_2^2-\frac{u_2^2}{\sinh^4\rho\sin^4\phi}),\label{eq:106}\\\notag\\\phi^{\prime\prime}=&\cos\phi\sin\phi(\omega_1^2-\frac{u_1^2}{\sinh^4\rho \cos^4\phi})-\sin\phi\cos\phi(\omega_2^2-\frac{u_2^2}{\sinh^4\rho\sin^4\phi})\notag\\&-2\phi^{\prime}\rho^{\prime}\coth\rho,\label{eq:107}
\end{align}
while the constraints give
\begin{align}
&\omega_1 u_1+\omega_2 u_2=\kappa u_3,\\\notag\\
&\rho^{\prime 2}+\phi^{\prime 2}\sinh^2\rho =\cosh^2\rho(\kappa^2+\frac{u_3^2}{\cosh^4\rho})- \sinh^2\rho\cos^2\phi(\omega_1^2+\frac{u_1^2}{\sinh^4\rho \cos^4\phi})\notag\\&-\sinh^2\rho\sin^2\phi(\omega_2^2+\frac{u_2^2}{\sinh^4\rho \sin^4\phi})\label{eq:105}
\end{align}
Notice that equation (\ref{eq:105}) 
is consistent with equation (\ref{eq:106}), for arbitrary constants $\kappa$, $\omega_1$,   $\omega_2$ and $u_i$, using also equation (\ref{eq:107}). 
\\\\ 
We now specialize to the case where $\omega_1=\omega_2=\omega$. Then equations (\ref{eq:106})-(\ref{eq:105}) lead to
\begin{align}
&\rho^{\prime \prime}=\cosh\rho\sinh\rho\bigg(\kappa^2-\omega^2+\phi^{\prime 2}-\frac{u_3^2}{\cosh^4\rho}+\frac{u_1^2}{\sinh^4\rho\cos^2\phi}+\frac{u_2^2}{\sinh^4\rho\sin^2\phi}\bigg)\label{new:9}
,\\\notag\\&\phi^{\prime\prime}=\cos\phi\sin\phi\left(\frac{u_2^2}{\sinh^4\rho\sin^4\phi}-\frac{u_1^2}{\sinh^4\rho \cos^4\phi}\right)-2\phi^{\prime}\rho^{\prime}\coth\rho,
\end{align}
while the constraints give
\begin{align}
&\omega (u_1+u_2)=\kappa u_3,\\\notag\\
\rho^{\prime 2}+\phi^{\prime 2}\sinh^2\rho =&\cosh^2\rho(\kappa^2+\frac{u_3^2}{\cosh^4\rho})- \omega^2\sinh^2\rho\notag\\&-\frac{u_1^2}{\sinh^2\rho \cos^2\phi}-\frac{u_2^2}{\sinh^2\rho \sin^2\phi}\label{new:7}
\end{align}
These four equations, equations \eqref{new:9}-\eqref{new:7} do not seem to be obviously solvable, unless $u_1=u_2=0$, but they are solvable. See the next Section. \\\\
Notice that for $u_i=0$ and $\phi^\prime=0$, then equations (\ref{new:9})-(\ref{new:7}) consistently lead to equation (\ref{ads3-2}), as we promised at the end of the previous section.

\section{Integrals of Motion for the two Spin String Solution in $AdS_5\times S^5$}
The three integrals of motion, which are in involuntion, are \\\\
\begin{align}
I_1=&-y_1^2+\frac{1}{\omega_1^2-\omega_2^2}\left[(y_1y_2^\prime-y_2y_1^\prime)^2+\frac{u_1^2}{y_1^2}y_2^2+\frac{u_2^2}{y_2^2}y_1^2\right]\notag\\&-\frac{1}{\omega_1^2-\kappa^2}\left[(y_1y_3^\prime-y_3y_1^\prime)^2+\frac{u_1^2}{y_1^2}y_3^2+\frac{u_3^2}{y_3^2}y_1^2\right]\\
I_2=&-y_2^2+\frac{1}{\omega_2^2-\omega_1^2}\left[(y_2y_1^\prime-y_1y_2^\prime)^2+\frac{u_2^2}{y_2^2}y_1^2+\frac{u_1^2}{y_1^2}y_2^2\right]\notag\\&-\frac{1}{\omega_2^2-\kappa^2}\left[(y_2y_3^\prime-y_3y_2^\prime)^2+\frac{u_2^2}{y_2^2}y_3^2+\frac{u_3^2}{y_3^2}y_2^2\right]\\
I_3=&y_3^2-\frac{1}{\kappa^2-\omega_1^2}\left[(y_3y_1^\prime-y_1y_3^\prime)^2+\frac{u_3^2}{y_3^2}y_1^2+\frac{u_1^2}{y_1^2}y_3^2\right]\notag\\&-\frac{1}{\kappa^2-\omega_2^2}\left[(y_3y_2^\prime-y_2y_3^\prime)^2+\frac{u_3^2}{y_3^2}y_2^2+\frac{u_2^2}{y_2^2}y_3^2\right]
\end{align}
Notice for consistency
\begin{equation}
I_1+I_2+I_3=1\label{new:8}
\end{equation}
Eventually, we will take $\omega_1=\omega_2=\omega$ and arbitrary $\kappa$. Obviously, the first non-trivial term in $I_1$ and in $I_2$ diverge for $\omega_1=\omega_2=\omega$, but we can still compute $I_1+I_2$ and $I_3$, which are finite.
\begin{align}
I_1+I_2 &=-\sinh^2\rho-\frac{\rho^{\prime 2}+\phi^{\prime 2}\sinh^2\rho\cosh^2\rho}{\omega^2-\kappa^2}\notag\\&-\frac{1}{\omega^2-\kappa^2}\bigg(\frac{u_3^2}{\cosh^2\rho}\sinh^2\rho+\left(\frac{u_1^2}{\sinh^2\rho\cos^2\phi}+\frac{u_2^2}{\sinh^2\rho\sin^2\phi}\right)\cosh^2\rho\bigg)\\
I_3&=\cosh^2\rho+\frac{\rho^{\prime 2}+\phi^{\prime 2} \sinh^2\rho\cosh^2\rho}{\omega^2-\kappa^2}\notag\\&+\frac{1}{\omega^2-\kappa^2}\bigg(\frac{u_3^2}{\cosh^2\rho}\sinh^2\rho+\left(\frac{u_1^2}{\sinh^2\rho\cos^2\phi}+\frac{u_2^2}{\sinh^2\rho\sin^2\phi}\right)\cosh^2\rho\bigg)
\end{align}
By using equation (\ref{new:7}), these two equations lead to    
\begin{align}
I_1+I_2&=-\frac{\kappa^{2}+\phi^{\prime 2}\sinh^{4}\rho+ u_3^{2}+u_1^{2}/ \cos^{2}\phi+u_2^{2}/ \sin^{2}\phi}{\omega^{2}-\kappa^{2}}\\
I_3&=\frac{\omega^{2}+\phi^{\prime 2}\sinh^{4}\rho+u_3^{2}+u_1^{2}/ \cos^{2}\phi+u_2^{2} / \sin^{2}\phi}{\omega^{2}-\kappa^{2}}
\end{align}
So that equation (\ref{new:8}) is trivially fulfilled. Without loss of generality, we can assume that $\omega^2>\kappa^2$. Actually we have to, to have periodic solutions, see \cite{Larsen:2003tb}. And furthermore, we can assume that both $I_1+I_2$ and $I_3$ are constants, as they should be. Then $I_3$ is a positive constant, say $C^2$, and $I_1+I_2$ is $1-C^2$
\begin{align}
I_1+I_2 = 1-C^2\label{new1:15}\\
I_3 = C^2\label{new1:16}
\end{align}
We now express equations \eqref{new:9}-\eqref{new:7} in terms of $C^2$. 
\begin{align}
&\rho^{\prime \prime}=\cosh\rho\sinh\rho\bigg(\kappa^2\left(1-\frac{C^2}{\sinh^4\rho}\right)+\omega^2\left(\frac{C^2-1}{\sinh^4\rho}-1\right)\notag\\&-u_3^2\left(\frac{1}{\cosh^4\rho}+\frac{1}{\sinh^4\rho}\right)\bigg),\label{new1:1}\\
&\phi^{\prime\prime}=\cos\phi\sin\phi\bigg(\notag\\&\bigg(\frac{\omega^{2}(C^2-1)}{\sinh^4\rho}-\frac{\kappa^{2}C^2}{\sinh^4\rho}-\phi^{\prime 2}-\frac{u_3^{2}}{\sinh^4\rho}\bigg)\left(\frac{1}{\sin^2\phi}-\frac{1}{\cos^2\phi}\right)\notag\\&+\frac{u_2^2-u_1^{2}}{\cos^{2}\phi\sinh^4\rho\sin^2\phi}\bigg)-2\phi^{\prime}\rho^{\prime}\coth\rho,\\
&\omega (u_1+u_2)=\kappa u_3,\\
&\rho^{\prime 2} =\kappa^2\left(\cosh^2\rho+\frac{C^2}{\sinh^2\rho}\right)+u_3^2\left(\frac{1}{\cosh^2\rho}+\frac{1}{\sinh^2\rho}\right)\notag\\&- \omega^2\left(\sinh^2\rho+\frac{C^2-1}{\sinh^2\rho}\right).\label{new1:2}
\end{align}
These four equations are obviously solvable, since equations \eqref{new1:1} and \eqref{new1:2} depend only on $\rho$. Notice also that these two equations are completely self-consistent, without using other equations. Finally, equation \eqref{new1:2}, is easily solved by substituting $y=\sinh^2\rho$. Then we get
\begin{align}
y^{\prime 2}&=4\bigg(\left(\kappa^2-  \omega^2\right)y^3+\left(2\kappa^2-\omega^2\right)y^2\notag\\&+\left(\kappa^2C^2+\kappa^2+2u_3^2-\omega^2C^2+\omega^2\right)y+\kappa^2C^2+u_3^2-\omega^2C^2+\omega^2\bigg)\label{new1:17}
\end{align}
This is a third order equation, which can be solved by an elliptic function.\\
We now specialize to the Neumann System, which means that $u_i=0$. Then there is a forth integral of motion, which is in involution with the three other integrals of motion in this sector:
\begin{align}
I=&\frac{1}{2}[-\omega^2_1(y^\prime_1)^2-\omega^2_2(y^\prime_2)^2+\kappa^2(y^\prime_3)^2\nonumber \\ &-((y^\prime_1)^2+(y^\prime_2)^2-(y^\prime_3)^2) 
(-\omega^2_1y^2_1-\omega^2_2y^2_2+\kappa^2y^2_3)-\omega^4_1y^2_1-\omega^4_2y^2_2+\kappa^4y^2_3]
\end{align}
For $u_i=0$, equations \eqref{new:9}-\eqref{new:7} consistently lead to
\begin{align}
&\phi^{\prime\prime}=-2\phi^{\prime}\rho^{\prime}\coth\rho,\label{eq:new30}\\
&\rho^{\prime 2}+\phi^{\prime 2}\sinh^2\rho =\kappa^2\cosh^2\rho- \omega^2\sinh^2\rho.
\end{align}
These two equations, trivially lead to
\begin{align}
&\phi^\prime = \frac{K}{\sinh^2\rho},\\
&\rho^{\prime 2} =\kappa^2\cosh^2\rho- \omega^2\sinh^2\rho-\frac{K^2}{\sinh^2\rho},
\end{align}
where $K$ is a constant. Then $C^2$ becomes
\begin{align}
C^2 = \frac{\omega^{2}+K^2}{\omega^{2}-\kappa^{2}}.\label{eq:new31}
\end{align}
Notice that these five equations, equations \eqref{eq:new30}-\eqref{eq:new31} are completely consistent with equations \eqref{new1:1}-\eqref{new1:2}, for $u_i=0$.
For the Neumann System, the final equation becomes
\begin{align}
y^{\prime 2}=4\bigg(\left(\kappa^2-  \omega^2\right)y^3+\left(2\kappa^2-\omega^2\right)y^2+\left(\kappa^2-K^2\right)y-K^2\bigg).\label{eq:new34}
\end{align}
Thus 
\begin{align}
y^\prime = 0,\quad \text{for } y=-1.
\end{align}
We further analyze equation \eqref{eq:new34}. Notice that it can be factorized as follows
\begin{align}
y^{\prime 2}=4 &\left(\kappa^{2}-\omega^{2}\right)(y+1)\notag \\
&\left(y-\left(\frac{\left(-\kappa^{2}+\sqrt{\kappa^{4}+4 K^{2}\left(\kappa^{2}-\omega^{2}\right)}\right)}{2\left(\kappa^{2}-\omega^{2}\right)}\right)\right.\notag\\
&\left(y-\left(\frac{\left(-\kappa^{2}-\sqrt{\kappa^{4}+4 K^{2}\left( \kappa^{2}-\omega^{2}\right)}\right)}{2\left( \kappa^{2}-\omega^{2}\right)}\right)\right).\label{new1:6}
\end{align}
To get periodic solutions, we are interested in the case with two non-negative roots, so we get the conditions 
\begin{align}
\omega^2>\kappa^2 \label{new1:13}\\
\kappa^{4}+4 K^{2}\left(\kappa^{2}-\omega^{2}\right) \geq 0
\end{align}
Equation \eqref{new1:6} is identical to equation (2.10) in reference \cite{Larsen:2003tb}, but in another notation, thus the solution is
\begin{align}
y^2(\sigma)=\frac{m}{n-m}\left(1-n \mathrm{sn}^{2}\left(\sqrt{\frac{b}{m}} \sigma \mid m\right)\right)
\end{align}
where \cite{Larsen:2003tb}
\begin{align}
&b= \sqrt{\kappa^{4}+4 K^{2}\left(\kappa^{2}-\omega^{2}\right)}\\
&m=\frac{2\sqrt{\kappa^{4}+4 K^{2}\left(\kappa^{2}-\omega^{2}\right)}}{2 \omega^{2}- \kappa^{2}+\sqrt{\kappa^{4}+4 K^{2}\left(\kappa^{2}-\omega^{2}\right)}}\\
&n=\frac{2 \sqrt{\kappa^{4}+4 K^{2}\left(\kappa^{2}-\omega^{2}\right)}}{\kappa^{2}+\sqrt{\kappa^{4}+4 K^{2}\left(\kappa^{2}-\omega^{2}\right)}}.
\end{align}
Notice also that \cite{Larsen:2003tb}
\begin{align}
b\geq 0,\quad 0,\leq m\leq n\leq 1
\end{align}
so that $y(\sigma)$ oscillates between the non-negative values $y_{\max }$ and $y_{\min }$, where
\begin{align}
y_{\max }=\sqrt{\frac{m}{n-m}}, \quad y_{\min }=\sqrt{\frac{m(1-n)}{n-m}}.
\end{align}
Notice that $y_{\max}$ can be very large, while $y_{\min}$ is still very small. So they don't necessarily follow each other, a comment to \cite{Tirziu:2009ed}.\\\\
Finally, the forth integral of motion leads to
\begin{align}
I=\frac{1}{2}\left[-\omega^2\kappa^2+(\omega^2-\kappa^2)K^2\right].
\end{align}
\section{Solving the Neumann-Rosochatius system on $AdS_3$}
\label{sec:9}
There is a standard procedure to solve the Neumann-Rosochatius integrable system on AdS. We first introduce the ellipsoidal coordinates $(\zeta_1, \zeta_2)$
\begin{align}
&y_1^2=\frac{(\omega_1^2-\zeta_1)(\omega_1^2-\zeta_2)}{\omega_{21}^2\omega_{13}^2},\quad y_2^2=\frac{(\omega_2^2-\zeta_1)(\omega_2^2-\zeta_2)}{\omega_{21}^2\omega_{32}^2},\quad y_3^2=\frac{(\omega_3^2-\zeta_1)(\omega_3^2-\zeta_2)}{\omega_{31}^2\omega_{32}^2},\label{new1:8}
\end{align}
where $\omega_{ij}^2=\omega_i^2-\omega_j^2\text{; } \omega_3=\kappa$. Notice that the $\omega_{ij}^2$ have their own algebra, for instance
\begin{align}
\omega_{12}^2+\omega_{23}^2+\omega_{31}^2=0,\quad \omega_{ij}^2=-\omega_{ji}^2.
\end{align}
Then it is almost trivial to show that equation (\ref{eq:71}) is fulfilled.\\\\
Actually, we have to be much more careful, since equation \eqref{new1:8} means that both $y_1^2$ and $y_2^2$ are ill-defined, since $\omega_{12}^2=-\omega_{21}^2$ is zero, for $\omega_1=\omega_2=\omega$. See Appendix A.\\\\
After some algebra, one finds the following separation equations
\begin{align}
\left(\frac{d\zeta_1}{d\sigma}\right)^2=-4\frac{P(\zeta_1)}{(\zeta_1-\zeta_2)^2},\quad \left(\frac{d\zeta_2}{d\sigma}\right)^2=-4\frac{P(\zeta_2)}{(\zeta_1-\zeta_2)^2},
\end{align}
where $P(\zeta)$ is
\begin{align}
P(\zeta)&=(\zeta-b_1)(\zeta-b_2)(\zeta-\omega^2)^2(\zeta-\kappa^2)\notag\\&+u_1^2(\zeta-\omega^2)^2(\zeta-\kappa^2)^2+u_2^2(\zeta-\omega^2)^2(\zeta-\kappa^2)^2+u_3^2(\zeta-\omega^2)^4.
\end{align}
where $b_1$ and $b_2$ are constants, which will be determined later.  Notice that $P(\zeta)$ can be written
\begin{align}
P(\zeta)=(\zeta-\omega^2)^2 P_1(\zeta)
\end{align}
where $P_1(\zeta)$ is
\begin{align}
P_1(\zeta)=(\zeta-b_1)(\zeta-b_2)(\zeta-\kappa^2)+u_1^2(\zeta-\kappa^2)^2+u_2^2(\zeta-\kappa^2)^2+u_3^2(\zeta-\omega^2)^2
\end{align}
Then we get 
\begin{align}
\frac{1}{(\zeta_1-\omega^2)^2}\left(\frac{d\zeta_1}{d\sigma}\right)^2&=-4\bigg(\frac{(\zeta_1-b_1)(\zeta_1-b_2)(\zeta_1-\kappa^2)}{(\zeta_1-\zeta_2)^2}\notag\\&+\frac{u_1^2(\zeta_1-\kappa^2)^2+u_2^2(\zeta_1-\kappa^2)^2+u_3^2(\zeta_1-\omega^2)^2}{(\zeta_1-\zeta_2)^2}\bigg),\\ \frac{1}{(\zeta_2-\omega^2)^2}\left(\frac{d\zeta_2}{d\sigma}\right)^2&=-4\bigg(\frac{(\zeta_2-b_1)(\zeta_2-b_2)(\zeta_2-\kappa^2)}{(\zeta_1-\zeta_2)^2}\notag\\&+\frac{u_1^2(\zeta_2-\kappa^2)^2+u_2^2(\zeta_2-\kappa^2)^2+u_3^2(\zeta_2-\omega^2)^2}{(\zeta_1-\zeta_2)^2}\bigg),
\end{align}
Mixing the two equations, we get
\begin{align}
&\left(\frac{d\zeta_1}{d\zeta_2}\right)^2=\frac{(\zeta_1-\omega^2)^2P_1(\zeta_1)}{(\zeta_2-\omega^2)^2P_1(\zeta_2)}
\end{align}
And separating the two variables, we get 
\begin{align}
&\int_{\zeta_{1_0}}^{\zeta_1}\frac{dy}{(y-\omega^2) \sqrt{P_1(y)}} = \pm\int_0^{\zeta_2}\frac{dy}{(y-\omega^2) \sqrt{P_1(y)}},\label{new:26}
\end{align}
where $\zeta_{1_0}$ is an integration constant. Notice that this equation gives a direct connection between $\zeta_1$ and $\zeta_2$, $\zeta_1(\zeta_2)$: See Appendix A. \\\\ We now specialize to the Neumann System again again. Then equation \eqref{new:26} leads to
\begin{align}
&\int_{\zeta_{1_0}}^{\zeta_1}\frac{dy}{(y-\omega^2) \sqrt{(y-b_1)(y-b_2)(y-\kappa^2)}}\notag\\& = \pm\int_0^{\zeta_2}\frac{dy}{(y-\omega^2) \sqrt{(y-b_1)(y-b_2)(y-\kappa^2)}}.\label{new1:4}
\end{align}
Following the procedure in reference \cite{Arutyunov:2003uj}, in details, we find the possibilities for $b_1$ and $b_2$
\begin{align}
&b_1 = \omega^{2}
+\kappa^{2}+K^2 &&b_2 = \omega^{2},\\ 
&b_1 = \omega^{2}  &&b_2 = \omega^{2}+\kappa^{2}+K^2.
\end{align}
In both cases, equation \eqref{new1:4} leads to
\begin{align}
&\int_{\zeta_{1_0}}^{\zeta_1}\frac{dy}{(y-\omega^2)^{\frac{3}{2}} \sqrt{(y-\kappa^2)(y-\omega^2-\kappa^2-K^2)}}\notag\\& = \pm\int_0^{\zeta_2}\frac{dy}{(y-\omega^2)^{\frac{3}{2}} \sqrt{(y-\kappa^2)(y-\omega^2-\kappa^2-K^2)}}.\label{new1:12}
\end{align}
This is an elliptic integral, which can be solved in a standard way, just like the elliptic equations \eqref{new1:17} and \eqref{eq:new34}. 
\section{The two Angular Momenta String Solution in $AdS_5\times S^5$}
We now start from scratch and take almost the opposite ansatz as in equation (\ref{eq:108}), i.e. constant "angels" in $AdS_5$. However, we have to take the time coordinate from $AdS_5$, so there is a little asymmetry between the two sectors. \\\\
The line element of $AdS_5\times S^5$ is \\
\begin{align}
ds^2=&-\cosh^2\rho dt^2 + d\rho^2 + \sinh^2\rho (d\phi^2 +\cos^2\phi d\phi_1^2+\sin^2\phi d\phi_2^2)\notag\\
& +d \theta^2+\cos^2 \theta d \psi_3^2+\sin^2 \theta(d \psi^2 +\cos^2\psi d \psi_1^2 +\sin^2 \psi d \psi_2^2)
\end{align}

\noindent The ansatz for the string solution in $AdS_5\times S^5$ is \begin{align}
&t=\kappa\tau\text{, }\rho=\rho_0\text{, }\phi=\phi_0\text{, }\phi_1=\phi_{1_0}\text{, }\phi_2=\phi_{2_0}\text{, }\notag\theta=\theta(\sigma)\text{, }\\&\psi=\psi(\sigma)\text{, } \psi_1=\omega_1\tau+\beta_1(\sigma)\text{, }\psi_2=\omega_2\tau+\beta_2(\sigma) \text{ and } \psi_3=\omega_3\tau+\beta_3(\sigma),
\end{align}
where $\rho_0$, $\phi_0$, $\phi_{1_0}$, $\phi_{2_0}$ are constants. \\\\
We introduce the embedding coordinates $(x_1,x_2,x_3)$
\begin{align}
x_1=\sin\theta\cos\psi,\quad x_2=\sin\theta\sin\psi,\quad x_3=\cos\theta
\end{align}
They fulfill the Neumann-Rosochatius  integrable system for the metric
\begin{equation}
h^{ij} = 
\begin{pmatrix}
1 & 0 & 0\\
0 & 1 & 0\\
0 & 0 & 1
\end{pmatrix}
\end{equation}
i.e.
\begin{gather}
h^{ij} x_i x_j\equiv x_1^2 +x_2^2+x_3^2=1\label{new:24}
\end{gather}
We have the periodicity conditions 
\begin{align}
x_i(\sigma+2\pi)=x_i(\sigma)\text{, }\quad \beta_i(\sigma+2\pi)=\beta(\sigma)+2\pi k_i,\label{eq(105)}
\end{align}
where $k_i$ are integers. Eventually, we will take $\omega_3=\kappa$, so eventually we will choose $k_3=0$.\\\\
The equations of motion determine that
\begin{align}
&\rho_0=0,\label{new:20}\\
&\theta^{\prime\prime}=\sin\theta\cos\theta(\psi^{\prime 2}+\omega_3^2-\beta_3^{\prime 2}-\cos^2\psi(\omega_1^2-\beta_1^{\prime 2})-\sin^2\psi(\omega_2^2-\beta_2^{\prime 2})),\\&\psi^{\prime\prime}=\cos\psi\sin\psi(\omega_1^2-\omega_2^2+\beta_2^{\prime 2}-\beta_1^{\prime 2})-2\psi^{\prime}\theta^{\prime}\cot\theta,\\&\beta_1^{\prime\prime}+2\beta_1^{\prime}(\theta^{\prime}\cot\theta-\psi^{\prime}\tan\psi)=0,\label{new:21}\\&\beta_2^{\prime\prime}+2\beta_2^{\prime}(\theta^{\prime}\cot\theta+\psi^{\prime}\cot\psi)=0,\label{new:22}\\&\beta_3^{\prime\prime}-2\beta_3^\prime\theta^{\prime}\tan\theta=0,\label{new:23}
\end{align}
while the constraint give
\begin{align}
&\omega_1\beta_1^{\prime}\sin^2\theta\cos^2\psi+\omega_2 \beta_2^{\prime}\sin^2\theta\sin^2\psi+\omega_3\beta_3^{\prime} \cos^2\theta =0\\\notag\\
&\theta^{\prime 2}+\psi^{\prime 2}\sin^2\theta+\sin^2\theta\cos^2\psi(\omega_1^2+\beta_1^{\prime 2})\notag\\&+\sin^2\theta\sin^2\psi(\omega_2^2+\beta_2^{\prime 2})+\cos^2\theta(\omega_3^2+\beta_3^{\prime 2})=\kappa^2
\end{align}
Notice that equation (\ref{new:20}) shows that there is a coordinate singularity at $\rho_0=0$, just like the event horizon is a coordinate singularity in the Schwarzschild black hole spacetime. But there is of course no curvature singularities in any of these cases. \\\\
Equations (\ref{new:21}), (\ref{new:22}) and (\ref{new:23}) are trivially solved by
\begin{align}
\beta_i^\prime = \frac{u_i}{x_i^2},\label{new:2}
\end{align}
where $u_i$ are constants. Then the remaining equations of motion lead to
\begin{align}
\theta^{\prime\prime}=&\sin\theta\cos\theta\bigg(\psi^{\prime 2}+\omega_3^2-\frac{u_3^2}{\cos^4\theta} -\cos^2\psi\left(\omega_1^2-\frac{u_1^2}{\sin^4\theta\cos^4\psi}\right)\notag\\&-\sin^2\psi\left(\omega_2^2-\frac{u_2^2}{\sin^4\theta\sin^4\psi}\right)\bigg),\label{eq:110}\\\psi^{\prime\prime}&=\cos\psi\sin\psi\left(\omega_1^2-\omega_2^2+\frac{u_2^2}{\sin^4\theta\sin^4\psi}-\frac{u_1^2}{\sin^4\theta\cos^4\psi}\right)-2\psi^{\prime}\theta^{\prime}\cot\theta \label{eq:111}
\end{align}
while the constraints give
\begin{align}
&\omega_1 u_1+\omega_2 u_2+\omega_3 u_3 =0\\\notag\\
&\theta^{\prime 2}+\psi^{\prime 2}\sin^2\theta=\kappa^2-\cos^2\theta\left(\omega_3^2+\frac{u_3^2}{\cos^4\theta}\right)-\sin^2\theta\cos^2\psi\left(\omega_1^2+\frac{u_1^2}{\sin^4\theta\cos^4\psi}\right)\notag\\&-\sin^2\theta\sin^2\psi\left(\omega_2^2+\frac{u_2^2}{\sin^4\theta\sin^4\psi}\right)\label{eq:109}
\end{align}
Notice that equation (\ref{eq:109}) is in some sense self-consistent, for arbitrary constants $\kappa$, $\omega_1$, $\omega_2$ and $u_i$, i.e. differentiating equation (\ref{eq:109}), we get an identity, using also equations (\ref{eq:110}) and (\ref{eq:111}).
\\\\ 
We now specialize to the case where $\omega_1=\omega_2=\omega$ and $\omega_3=\kappa$. Then equations (\ref{eq:110})-(\ref{eq:109}) lead to 
\begin{align}
\theta^{\prime\prime}&=\sin\theta\cos\theta\left(\psi^{\prime 2}+\kappa^2-\omega^2 +\frac{u_1^2}{\sin^4\theta\cos^2\psi}+\frac{u_2^2}{\sin^4\theta\sin^2\psi}-\frac{u_3^2}{\cos^4\theta}\right)\label{new:27},\\\psi^{\prime\prime}&=\cos\psi\sin\psi\left(\frac{u_2^2}{\sin^4\theta\sin^4\psi}-\frac{u_1^2}{\sin^4\theta\cos^4\psi}\right)-2\psi^{\prime}\theta^{\prime}\cot\theta 
\end{align}
while the constraints give
\begin{align}
&\omega (u_1+ u_2)+\kappa u_3 =0\\\notag\\
&\theta^{\prime 2}+\psi^{\prime 2}\sin^2\theta=(\kappa^2-\omega^2)\sin^2\theta-\frac{u_3^2}{\cos^2\theta}-\frac{u_1^2}{\sin^2\theta\cos^2\psi}-\frac{u_2^2}{\sin^2\theta\sin^2\psi}\label{new:14}
\end{align}
These four equations, equations \eqref{new:27}-\eqref{new:14} do not seem to be obviously solvable, unless $u_1=u_2=0$, but they are solvable. See the next Section.

\section{Integrals of Motion for the two Angular Momenta String Slution in $AdS_5\times S^5$}
The three integrals of motion, which are in involuntion, are 

\begin{align}
F_1=&x_1^2+\frac{1}{\omega_1^2-\omega_2^2}\left[(x_1x_2^\prime-x_2x_1^\prime)^2+\frac{u_1^2}{x_1^2}x_2^2+\frac{u_2^2}{x_2^2}x_1^2\right]\notag\\&+\frac{1}{\omega_1^2-\omega^2_3}\left[(x_1x_3^\prime-x_3x_1^\prime)^2+\frac{u_1^2}{x_1^2}x_3^2+\frac{u_3^2}{x_3^2}x_1^2\right]\\
F_2=&x_2^2+\frac{1}{\omega_2^2-\omega_1^2}\left[(x_2x_1^\prime-x_1x_2^\prime)^2+\frac{u_2^2}{x_2^2}x_1^2+\frac{u_1^2}{x_1^2}x_2^2\right]\notag\\&+\frac{1}{\omega_2^2-\omega_3^2}\left[(x_2x_3^\prime-x_3x_2^\prime)^2+\frac{u_2^2}{x_2^2}x_3^2+\frac{u_3^2}{x_3^2}x_2^2\right]\\
F_3=&x_3^2+\frac{1}{\omega_3^2-\omega_1^2}\left[(x_3x_1^\prime-x_1x_3^\prime)^2+\frac{u_3^2}{x_3^2}x_1^2+\frac{u_1^2}{x_1^2}x_3^2\right]\notag\\&+\frac{1}{\omega_3^2-\omega_2^2}\left[(x_3x_2^\prime-x_2x_3^\prime)^2+\frac{u_3^2}{x_3^2}x_2^2+\frac{u_2^2}{x_2^2}x_3^2\right]
\end{align}
Notice for consistency
\begin{align}
F_1+F_2+F_3=1\label{new:10}
\end{align}
Eventually, we will take $\omega_1=\omega_2=\omega$ and $\omega_3=\kappa$.
Obviously, the first non-trivial term in $F_1$ and in $F_2$ diverge for $\omega_1=\omega_2=\omega$, but we can still compute $F_1+F_2$ and $F_3$, which are finite.
\begin{align}
F_1+F_2&=\sin^2\theta+ \frac{\theta^{\prime2}+\psi^{\prime 2}\sin^2\theta \cos^2\theta}{\omega^2-\kappa^2}\notag\\&+\frac{1}{\omega^2-\kappa^2}\bigg[\frac{u_3^2}{\cos^2\theta}\sin^2\theta+\left(\frac{u_1^2}{\sin^2\theta\cos^2\psi}+\frac{u_2^2}{\sin^2\theta\sin^2\psi}\right)\cos^2\theta\bigg]\\
F_3&=\cos^2\theta-\frac{\theta^{\prime2}+\psi^{\prime 2}\sin^2\theta \cos^2\theta}{\omega^2-\kappa^2}\notag\\&- \frac{1}{\omega^2-\kappa^2}\bigg[\frac{u_3^2}{\cos^2\theta}\sin^2\theta+\left(\frac{u_1^2}{\sin^2\theta\cos^2\psi}+\frac{u_2^2}{\sin^2\theta\sin^2\psi}\right)\cos^2\theta\bigg]
\end{align}
By using equation (\ref{new:14}), these two equations lead to    
\begin{align}
&F_1+F_2=-\frac{\psi'^{2}\sin^{4}\theta+u_3^{2}+u_1^{2}/\cos^{2}\psi+u_2^{2}/\sin^{2}\psi}{\omega^{2}-\kappa^{2}}\\&
F_3=1+\frac{\psi'^{2}\sin^{4}\theta+u_3^{2}+u_1^{2}/\cos^{2}\psi+u_2^{2}/\sin^{2}\psi}{\omega^{2}-\kappa^{2}}
\end{align}
So that equation (\ref{new:10}) is trivially fulfilled. Without loss of generality, we can assume that $\kappa^{2}>\omega^2$. Actually we have to, to have periodic solutions, see later in this Section. And furthermore, we can assume that both $F_1+F_2$ and $F_3$ are constants, as they should be. Then $F_1+F_2$ is a positive constant, say $C^2$, and $F_3$ is $1-C^2$
\begin{align} 
&F_1+F_2=C^2\label{new:12}\\
&F_3=1-C^2\label{new:11}
\end{align}
We now express equations \eqref{new:27}-\eqref{new:14} in terms of $C^2$. 
\begin{align}
\theta^{\prime\prime}&=\sin\theta\cos\theta\bigg(\kappa^2\left(1+\frac{C^2}{\sin^4\theta}\right)-\omega^2\left(1+\frac{C^2}{\sin^4\theta}\right) -u_3^{2}\left(\frac{1}{\sin^4\theta}+\frac{1}{\cos^4\theta}\right)\bigg),\label{new:28}\\
\psi^{\prime\prime}&=\cos\psi\sin\psi\bigg(\left(\psi'^{2}+\frac{u_3^{2}}{\sin^4\theta}+\frac{\omega^2C^2}{\sin^4\theta}-\frac{\kappa^2C^2}{\sin^4\theta}\right)\left(\frac{1}{\cos^2\psi}-\frac{1}{\sin^2\psi}\right)\notag\\&+\frac{u_2^{2}-u_1^{2}}{\sin^{2}\psi\sin^4\theta\cos^2\psi}\bigg)-2\psi^{\prime}\theta^{\prime}\cot\theta 
\end{align}
while the constraints give
\begin{align}
&\omega (u_1+ u_2)+\kappa u_3 =0\\\notag\\
&\theta^{\prime 2}=\kappa^2\left(\sin^2\theta-\frac{C^2}{\sin^2\theta}\right)-\omega^2 \left(\sin^2\theta-\frac{C^2}{\sin^2\theta}\right)+u_3^2\left(\frac{1}{\sin^2\theta}-\frac{1}{\cos^2\theta}\right)\label{new:29}
\end{align}
These four equations are not obviously solvable, since equations \eqref{new:28} and \eqref{new:29} depend only on $\theta$. Notice also that these two equations are completely self-consistent, without using other equations. Finally, equation \eqref{new:29}, is easily solved by substituting $x=\sin^2\theta$. Then we get
\begin{align}
x^{\prime 2}=&4\bigg(\left(\omega^2-\kappa^2\right)x^3+\left(\kappa^2- \omega^2\right)x^2+\left(C^2\left(\kappa^2-\omega^2\right)-2u_3^2\right)x\notag\\&+C^2\left(\omega^2-\kappa^2\right)+u_3^2\bigg)\label{new1:18}
\end{align}
This is a third order equation, which can be solved by an elliptic function.
We now specialize to the Neumann System, which means that $u_i=0$. Then there is a fourth integral of motion, which is in involution with the three other integrals of motion in this sector:
\begin{align}
F=&\frac{1}{2}[\omega^2_1(x^\prime_1)^2+\omega^2_2(x^\prime_2)^2+\omega_3^2(x^\prime_3)^2\nonumber \\ &+((x^\prime_1)^2+(x^\prime_2)^2+(x^\prime_3)^2) 
(\omega^2_1x^2_1+\omega^2_2x^2_2+\omega_3^2x^2_3)+\omega^4_1x^2_1+\omega^4_2x^2_2+\omega_3^4x^2_3]
\end{align}
For $u_i=0$, equations \eqref{new:27}-\eqref{new:14} consistently lead to
\begin{align}
&\psi^{\prime\prime}=-2\psi^{\prime}\theta^{\prime}\cot\theta\label{eq:new32} \\
&\theta^{\prime 2}+\psi^{\prime 2}\sin^2\theta=(\kappa^2-\omega^2)\sin^2\theta
\end{align}
These two equations, trivially lead to
\begin{align}
&\psi^\prime = \frac{K}{\sin^2 \theta}\\
&\theta^{\prime 2}=(\kappa^2-\omega^2)\sin^2\theta-\frac{K^2}{\sin^2 \theta},
\end{align}
where $K$ is a constant. Then $C^2$ becomes
\begin{align}
C^2 = \frac{K^2}{\kappa^{2}-\omega^{2}}.\label{eq:new33}
\end{align}
Notice that these five equations, equations \eqref{eq:new32}-\eqref{eq:new33} are completely consistent with equations \eqref{new:28}-\eqref{new:29}, for $u_i=0$. 
For the Neumann System, the final equation becomes
\begin{align}
x^{\prime 2}=4\bigg(\left(\omega^2-\kappa^2\right)x^3+\left(\kappa^2- \omega^2\right)x^2+K^2x- K^2\bigg)\label{new1:3}
\end{align}
Thus
\begin{align}
x^\prime = 0,\quad \text{for } x=1.
\end{align}
We further analyze equation \eqref{new1:3}. Notice that it can be factorized as follows
\begin{align}
x^{\prime 2}=4\left(\omega^2-\kappa^2\right)(x-1)\left(x-\frac{K}{\sqrt{\kappa^2-\omega^2}}\right)\left(x+\frac{K}{\sqrt{\kappa^2-\omega^2}}\right)
\end{align}
To get periodic solutions, we are interested in the case with two non-negative roots, so we get the condition
\begin{align}
\kappa^2>\omega^2.\label{new1:14}
\end{align}
We now follow the procedure of \cite{Larsen:2003tb}, and define the function $P$
\begin{align}
x = \frac{P}{\omega^2-\kappa^2}+\frac{1}{3}
\end{align}
such that
\begin{align}
P^{\prime 2} &= 4 P^3-g_2P-g_3\notag\\
&=4(P-e_1)(P-e_2)(P-e_3)\label{new1:11}
\end{align}
where the roots $(e_1>e_2\geq e_3)$ are 
\begin{align}
&e_1 =\frac{\kappa^2-\omega^2}{3}+ K \sqrt{\kappa^2-\omega^2},\\&e_2 = \frac{\kappa^2-\omega^2}{3}- K \sqrt{\kappa^2-\omega^2},\\&e_3 = \frac{2(\omega^2-\kappa^2)}{3}.
\end{align}
The invariants $g_2=2(e_1^2+e_2^2+e_3^2)$ and $g_3=4e_1e_2e_3$. Notice also that $\Delta \equiv g_2^3-27g_3^2\geq 0$, using \eqref{new1:11}. Equation \eqref{new1:11} is the Weierstrass equation with solution
\begin{align}
P(\sigma)=\wp(\sigma+a)
\end{align}
where $\wp$ is the doubly periodic Weierstrass function and $a$ is a complex integration constant \cite{Abramowitz72}. Following the procedure of \cite{Ryang:2004tq}, we find that $x=\cos\theta$ can be expressed in terms of the Jacobi elliptic function $\operatorname{dc}(u,m)$

\begin{align}
&x(\sigma)=\sqrt{a_+} \operatorname{dc}(\sqrt{a_+(\kappa^2-\omega^2)}\sigma ,m)\text{, where}\\
&a_\pm = \frac{\kappa^2-\omega^2\pm K\sqrt{\left(\kappa^{2}-\omega^2\right)}}{\kappa^2-\omega^2}\\
&m = \frac{a_-}{a_+}= \frac{\kappa^2-\omega^2- K\sqrt{\left(\kappa^{2}-\omega^2\right)}}{\kappa^2-\omega^2+ K\sqrt{\left(\kappa^{2}-\omega^2\right)}}.
\end{align}
Finally, we get that
\begin{align}
x^2(\sigma) = a_+ \operatorname{dc}^2(\sqrt{a_+(\kappa^2-\omega^2)}\sigma ,m).
\end{align}
This result is, in some sense, conjugate to the result in equation \eqref{new1:3}, since there we used $x=\sin^2\theta$, but here we use $x=\cos\theta$.\\\\ 
Finally, the forth integral of motion leads to
\begin{align}
F =\frac{1}{2}[
\kappa^4-(\kappa^2-\omega^2)K^2 ].
\end{align}
\section{Solving the Neumann-Rosochatius system on $S^3$}
There is a standard procedure to solve the Neumann-Rosochatius integrable system on the sphere. We first introduce the ellipsoidal coordinates $(\zeta_1,\zeta_2)$
\begin{align}
x_1^2=\frac{(\omega_1^2-\zeta_1)(\omega_1^2-\zeta_2)}{\omega_{12}^2\omega_{13}^2},\quad x_2^2=\frac{(\omega_2^2-\zeta_1)(\omega_2^2-\zeta_2)}{\omega_{21}^2\omega_{23}^2},\quad x_3^2=\frac{(\omega_3^2-\zeta_1)(\omega_3^2-\zeta_2)}{\omega_{31}^2\omega_{32}^2},\label{new1:9}
\end{align}
where $\omega_{ij}^2=\omega_i^2-\omega_j^2; \omega_3=\kappa$. Notice that the $\omega_{ij}^2$ have their own algebra, for instance 
\begin{align}
\omega_{12}^2+\omega_{23}^2+\omega_{31}^2=0,\quad \omega_{ij}^2=-\omega_{ji}^2.
\end{align} Then it is almost trivial to show that equation (\ref{new:24}) is fulfilled.\\\\
Actually, we have to be much more careful, since equation \eqref{new1:9} means that both $x_1^2$ and $x_2^2$ are ill-defined, since $\omega_{12}^2=-\omega_{21}^2$ is zero, for $\omega_1=\omega_2=\omega$. See Appendix A.\\\\
After some algebra, one finds the following separation equations
\begin{align}
\left(\frac{d\zeta_1}{d\sigma}\right)^2=-4\frac{P(\zeta_1)}{(\zeta_1-\zeta_2)^2},\quad \left(\frac{d\zeta_2}{d\sigma}\right)^2=-4\frac{P(\zeta_2)}{(\zeta_1-\zeta_2)^2},
\end{align}
where $P(\zeta)$ is
\begin{align}
P(\zeta)&=(\zeta-b_1)(\zeta-b_2)(\zeta-\omega^2)^2(\zeta-\kappa^2)\notag\\&+u_1^2(\zeta-\omega^2)^2(\zeta-\kappa^2)^2+u_2^2(\zeta-\omega^2)^2(\zeta-\kappa^2)^2+u_3^2(\zeta-\omega^2)^4.
\end{align}
where $b_1$ and $b_2$ are constants, which will be determined later.  Notice that $P(\zeta)$ can be written
\begin{align}
P(\zeta)=(\zeta-\omega^2)^2P_1(\zeta)
\end{align}
where $P_1(\zeta)$ is
\begin{align}
P_1(\zeta)&=(\zeta-b_1)(\zeta-b_2)(\zeta-\kappa^2)+u_1^2(\zeta-\kappa^2)^2+u_2^2(\zeta-\kappa^2)^2+u_3^2(\zeta-\omega^2)^2
\end{align}
Then we get 
\begin{align}
\frac{1}{(\zeta_1-\omega^2)^2}\left(\frac{d\zeta_1}{d\sigma}\right)^2&=-4\bigg(\frac{(\zeta_1-b_1)(\zeta_1-b_2)(\zeta_1-\kappa^2)}{(\zeta_1-\zeta_2)^2}\notag\\&+\frac{u_1^2(\zeta_1-\kappa^2)^2+u_2^2(\zeta_1-\kappa^2)^2+u_3^2(\zeta_1-\omega^2)^2}{(\zeta_1-\zeta_2)^2}\bigg),\\
\frac{1}{(\zeta_2-\omega^2)^2}\left(\frac{d\zeta_2}{d\sigma}\right)^2&=-4\bigg(\frac{(\zeta_2-b_1)(\zeta_2-b_2)(\zeta_2-\kappa^2)}{(\zeta_1-\zeta_2)^2}\notag\\&+\frac{u_1^2(\zeta_2-\kappa^2)^2+u_2^2(\zeta_2-\kappa^2)^2+u_3^2(\zeta_2-\omega^2)^2}{(\zeta_1-\zeta_2)^2}\bigg),
\end{align}
Mixing the two equations, we get 
\begin{align}
&\left(\frac{d\zeta_1}{d\zeta_2}\right)^2=\frac{(\zeta_1-\omega^2)^2P_1(\zeta_1)}{(\zeta_2-\omega^2)^2P_1(\zeta_2)}
\end{align}
And separating the two variables, we get 
\begin{align}
&\int_{\zeta_{1_0}}^{\zeta_1}\frac{dx}{(x-\omega^2) \sqrt{P_1(x)}} =\pm \int_0^{\zeta_2}\frac{dx}{(x-\omega^2) \sqrt{P_1(x)}},\label{new:25}
\end{align}
where $\zeta_{1_0}$ is an integration constant. Notice that this equation gives a direct connection between $\zeta_1$ and $\zeta_2$, $\zeta_1(\zeta_2)$: See Appendix A. \\\\ We now specialize to the Neumann System again again. Then equation \eqref{new:25} leads to
\begin{align}
&\int_{\zeta_{1_0}}^{\zeta_1}\frac{dx}{(x-\omega^2) \sqrt{(x-b_1)(x-b_2)(x-\kappa^2)}}\notag\\& =\pm \int_0^{\zeta_2}\frac{dx}{(x-\omega^2) \sqrt{(x-b_1)(x-b_2)(x-\kappa^2)}}.\label{new1:5}
\end{align}
Following the procedure in reference \cite{Arutyunov:2003uj}, in details, we find the possibilities for $b_1$ and $b_2$
\begin{align}
&b_1 = K^2+\omega^2, &&b_2 = \omega^{2}\quad \\ 
&b_1 = \omega^{2},  &&b_2 = K^2+\omega^2.
\end{align}
In both cases, equation \eqref{new1:5} leads to
\begin{align}
&\int_{\zeta_{1_0}}^{\zeta_1}\frac{dx}{(x-\omega^2)^{\frac{3}{2}} \sqrt{(x-\kappa^2)(x-K^2-\omega^2)}}\notag\\& =\pm \int_0^{\zeta_2}\frac{dx}{(x-\omega^2)^{\frac{3}{2}} \sqrt{(x-\kappa^2)(x-K^2-\omega^2)}}.\label{new1:19}
\end{align}
This is an elliptic integral, which can be solved in a standard way, just like the elliptic equations \eqref{new1:18} and \eqref{new1:3}.
\section{Preliminary Conclusions}
In this paper, we have shown that two almost opposite sectors of string configuration ansätze in $AdS_5\times S^5$, anyway have given almost the same results\footnote{Actually, I dreamt these results beforehand, probably in 2016. I always dream my important results beforehand. In the same dream, I dreamt that there were lions in the woods of Denmark....}. Or in some sense, the opposite results, compare equation \eqref{new1:13}  and equation \eqref{new1:14}. Compare also equations \eqref{new1:15}-\eqref{new1:16} and equations \eqref{new:12}-\eqref{new:11}. We want, first of all, to generalize these results to the five dimensional $AdS$ black hole \cite{Larsen:2007va} $\times S^5$. The five dimensional $AdS$ black hole has some of the same properties as $AdS_5$ \cite{Larsen:2003tb}. We want, of course, also to generalize these results to the Neumann-Rosochatius System \cite{Arutyunov:2003za, ADLER1980318, 10.1007/978-1-4613-8109-9_7, Ratiu1981EulerPoissonEO, Ratiu1981TheCN, Reiman82, RosochatiusPHD, Chakraborty:2019gmt, Ahn:2008hj, Hernandez:2015nba, Chakraborty:2020las} and to several types of deformed Neumann-Rosochatius Systems \cite{Hernandez:2017raj, Arutyunov:2016ysi, Nieto:2017wtp}. It would be interesting to investigate whether all these string solutions are stable under progagation of string perturbations \cite{Larsen:1993iva}?\\\\
We now turn to the second part of this paper, which is independent of any string configurations in $D=10$ String Theory and independent of any configurations in $D=11$ M-theory. This part concerns String Cosmology: 

\section{True Conclusions}
In $D=10$ String Cosmology there is no Big Bang \cite{gasperini_2007}, cf. the work of Maurizio Gasperini, Gabriele Veneziano, Massimo Giovannini etc. Yes, indeed, there is a relatively big Big Bang, but there is a pre Big Bang and there is a continuous transition through the relatively big Big Bang, and there is Inflation, Cosmic Microwave Background Radiation and all the usual stuff in the early Universe, but there is no true Singularity. So there is no boundary conditions for the wave function of the Universe at the origin of Time, because there were no origin of Time. That there is no Big Bang is completely consistent with, for instance \cite{deVega:2001wq}\footnote{In my first year of ALS, I could hardly walk, I went to Kharkiv in Ukraine, to give a talk about this work. Now, my only wish is peace and freedom for the Ukranian people.}.\\
\indent Listen carefully to me, because I will say this only once, I seriously mean that Stephen Hawking and Roger Penrose \cite{Hawking1970TheSO} were wrong about the Big Bang, I truly believe that the Universe has been there forever. \\
\indent Many people have considered $dS_4$ as a part of String Cosmology in $D=10$ dimensions, which is somewhat in accordance with our observations in four dimensions. In String Cosmology, it should of course be something like $dS_4\times T^4\times S^2$ \cite{Obied:2018sgi, Agrawal:2018own, Berglund:2019pxr}, cf. the work of Cumrun Vafa, Hirosi Ooguri, Per Berglund etc. In $dS_4$, there is of course a positive cosmological constant, or an effective positive cosmological constant. In this case, there is an effective positive cosmological constant, due to dark energy \cite{Berglund:2019ctg}. In this case, the dark energy comes from the axion-dilaton background and personally, I don't believe in dark matter. String Theories give a renormalizable correction to General Relativity, and there should be a way out of the flat galaxy rotation curves? The four dimensional de Sitter space, $dS_4$, should of course be a somewhat asymmetrical time-dependent de Sitter space, since there were no energy in the beginning. A somewhat asymmetrical time-dependent de Sitter space is crucial, to create matter.\\
\indent Listen carefully to me, because I will say this only once, I seriously mean that Stephen Hawking and Roger Penrose \cite{Hawking1970TheSO} were wrong about the Big Bang, I truly believe that the Universe has been....\\\\
POST SCRIPTUM: In a Universe where you can travel in time, back and forwards, the present time depends both on the past time, within the light cone, and the future time, within the light cone, cf. Novikov's self-consistency conjecture, so that physics is consistent. This is well known by time travelers, including me. So why bother travel back in time?

\begin{center}
ACKNOWLEDGMENTS:
\end{center}
I would like to thank my scientific secretary, Daniel N. Iversen, for tremendous secretarial work and for doing all my trivial computations, actually we did them together\footnote{See the footnote at page 1. When Stephen Hawking still was alive, i therefore always called him for a sissy, since it is only sissies that can breathe themselves...}.
\vspace{3cm}
\section*{Appendix A}
Notice that the denominators in the two first terms in equation \eqref{new1:8} give zero, in the limit where $\omega_1=\omega_2=\omega$. Therefore, the numerators in the two first terms, in the same limit, must also give zero. \\
\indent By differentiating equation \eqref{new1:12}, we find that the relationship between $\zeta_1$ and $\zeta_2$ can be at most linear
\begin{align}
\zeta_1 = \alpha \zeta_2+\beta,
\end{align}
where $\alpha$ and $\beta$ are constants.\\
\indent Inserting this in the two first terms in equation \eqref{new1:8}, we get
\begin{align}
\alpha \zeta_2^2+(\beta-\alpha \omega^2-\omega^2)\zeta_2+\omega^4-\beta\omega^2=0
\end{align}
to have zero in the numerators. And we find exactly the same relationship in equation \eqref{new1:19}.

\newpage
\bibliographystyle{ieeetr}
\bibliography{INSPIRE-CiteAll}

\end{document}